\documentclass[sn-basic]{sn-jnl}% Basic Springer Nature Reference Style
\usepackage{graphicx}%
\usepackage{multirow}%
\usepackage{amsmath,amssymb,amsfonts}%
\usepackage{amsthm}%
\usepackage{mathrsfs}%
\usepackage[title]{appendix}%
\usepackage{xcolor}%
\usepackage{textcomp}%
\usepackage{manyfoot}%
\usepackage{booktabs}%
\usepackage{algorithm}%
\usepackage{algorithmicx}%
\usepackage{algpseudocode}%
\usepackage{listings}%
\newcommand{\mps}{\textup{m/s}}
\newcommand{\cmps}{\textup{cm/s}}

\begin{document}

\title[Cosmology and fundamental physics with ANDES]{Cosmology and fundamental physics with the ELT-ANDES spectrograph}

\author*[1,2]{\fnm{C.J.A.P.} \sur{Martins}}\email{Carlos.Martins@astro.up.pt}
\author[3]{\fnm{R.} \sur{Cooke}}
\author[4]{\fnm{J.} \sur{Liske}}
\author[5]{\fnm{M.T.} \sur{Murphy}}
\author[6,7]{\fnm{P.} \sur{Noterdaeme}}
\author[8]{\fnm{T.M.} \sur{Schmidt}}
\author[9]{\fnm{J.S.} \sur{Alcaniz}}
\author[1,10]{\fnm{C.S.} \sur{Alves}}
\author[11]{\fnm{S.} \sur{Balashev}}
\author[12,13,14]{\fnm{S.} \sur{Cristiani}}
\author[12]{\fnm{P.} \sur{Di Marcantonio}}
\author[15,16]{\fnm{R.} \sur{G\'enova Santos}}
\author[17,9]{\fnm{R.S.} \sur{Gon\c calves}}
\author[15,16]{\fnm{J. I.} \sur{Gonz\'alez Hern\'andez}}
\author[18]{\fnm{R.} \sur{Maiolino}}
\author[19]{\fnm{A.} \sur{Marconi}}
\author[1,2,20]{\fnm{C.M.J.} \sur{Marques}}
\author[1,20]{\fnm{M.A.F.} \sur{Melo e Sousa}}
\author[21]{\fnm{N.J.} \sur{Nunes}}
\author[22]{\fnm{L.} \sur{Origlia}}
\author[23,24]{\fnm{C.} \sur{P\'eroux}}
\author[25]{\fnm{S.} \sur{Vinzl}}
\author[26]{\fnm{A.} \sur{Zanutta}}
\affil[1]{\orgdiv{Centro de Astrof\'{\i}sica da Universidade do Porto}, \orgaddress{\street{Rua das Estrelas}, \city{4150-762 Porto}, \country{Portugal}}}
\affil[2]{\orgdiv{Instituto de Astrof\'{\i}sica e Ci\^encias do Espa\c co}, \orgname{Universidade do Porto}, \orgaddress{\street{Rua das Estrelas}, \city{4150-762 Porto}, \country{Portugal}}}
\affil[3]{\orgdiv{Centre for Extragalactic Astronomy, Durham University}, \orgaddress{\street{Science Site, South Road}, \city{DH1 3LE, Durham}, \country{UK}}}
\affil[4]{\orgdiv{Hamburger Sternwarte}, \orgname{Universit{\"a}t Hamburg}, \orgaddress{\street{Gojenbergsweg 112}, \city{21029 Hamburg}, \country{Germany}}}
\affil[5]{Centre for Astrophysics and Supercomputing, Swinburne University of Technology, Hawthorn, Victoria 3122, Australia}
\affil[6]{\orgdiv{Institut d'Astrophysique de Paris, UMR 7095, CNRS-SU}, \orgaddress{\street{98bis bd Arago}, \city{75014 Paris}, \country{France}}}
\affil[7]{\orgdiv{French-Chilean Laboratory for Astronomy, IRL 3386, CNRS and U. de Chile}, \orgaddress{\street{Casilla 36-D}, \city{Santiago}, \country{Chile}}}
\affil[8]{Observatoire Astronomique de l'Universit\'e de Gen\`eve, Chemin Pegasi 51, Sauverny, CH-1290, Switzerland}
\affil[9]{Observat\'orio Nacional, 20921-400, Rio de Janeiro, RJ, Brazil}
\affil[10]{Department of Physics and Astronomy, University College London, Gower Street, London WC1E 6BT, United Kingdom}
\affil[11]{Ioffe Institute, {Polyteknicheskaya 26}, 194021 Saint-Petersburg, Russia}
\affil[12]{INAF--Osservatorio Astronomico di Trieste, Via G.B. Tiepolo, 11, I-34143 Trieste, Italy}
\affil[13]{IFPU--Institute for Fundamental Physics of the Universe, via Beirut 2, I-34151 Trieste, Italy}
\affil[14]{INFN-National Institute for Nuclear Physics, via Valerio 2, I-34127 Trieste}
\affil[15]{Instituto de Astrof{\'\i}sica de Canarias, E-38205 La Laguna, Tenerife, Spain
}
\affil[16]{Universidad de La Laguna, Dept. Astrof{\'\i}sica, E-38206 La Laguna, Tenerife, Spain}
\affil[17]{Departamento de F\'isica, Universidade Federal Rural do Rio de Janeiro, 23897-000, Serop\'edica, RJ, Brazil}
\affil[18]{Cavendish Laboratory, University of Cambridge, 19 J.J. Thomson Ave., Cambridge CB3 0HE, UK}
\affil[19]{Dipartimento di Fisica e Astronomia, Universita\`a degli Studi di Firenze, Via G. Sansone 1, I-50019, Sesto Fiorentino, Firenze, Italy}
\affil[20]{Faculdade de Ci\^encias, Universidade do Porto, Rua do Campo Alegre, 4150-007 Porto, Portugal}
\affil[32]{Instituto de Astrof\'isica e Ci\^encias do Espa\c{c}o, Faculdade de Ci\^encias da Universidade de Lisboa, Edif\'icio C8, Campo Grande, 1749-016 Lisboa}
\affil[22]{INAF--Osservatorio di Astrofisica e Scienza dello Spazio di Bologna, Via Gobetti 93/3, I-40129 Bologna, Italy}
\affil[23]{European Southern Observatory, Karl-Schwarzschild-Str. 2, 85748 Garching-bei-M\"unchen, Germany}
\affil[24]{Aix Marseille Universit\'e, CNRS, LAM (Laboratoire d'Astrophysique de Marseille) UMR 7326, 13388, Marseille, France}
\affil[25]{Universit\'e de Toulouse UPS, Toulouse, France}
\affil[26]{INAF--Osservatorio Astronomico di Brera, via E. Bianchi 46, 23807 Merate, Italy}

\abstract{State-of-the-art 19th century spectroscopy led to the discovery of quantum mechanics, and 20th century spectroscopy led to the confirmation of quantum electrodynamics. State-of-the-art 21st century astrophysical spectrographs, especially ANDES at ESO's ELT, have another opportunity to play a key role in the search for, and characterization of, the new physics which is known to be out there, waiting to be discovered. We rely on detailed simulations and forecast techniques to discuss four important examples of this point: big bang nucleosynthesis, the evolution of the cosmic microwave background temperature, tests of the universality of physical laws, and a real-time model-independent mapping of the expansion history of the universe (also known as the redshift drift). The last two are among the flagship science drivers for the ELT. We also highlight what is required for the ESO community to be able to play a meaningful role in 2030s fundamental cosmology and show that, even if ANDES only provides null results, such `minimum guaranteed science' will be in the form of constraints on key cosmological paradigms: these are independent from, and can be competitive with, those obtained from traditional cosmological probes.}

\keywords{Cosmology, Fundamental Physics, High-resolution spectroscopy, ANDES}
\maketitle

\section{Introduction}

State-of-the-art spectroscopy has been crucial to the development of our contemporary fundamental physics paradigm. It has been a key motivation for the development of quantum mechanics (cf. the discrete nature of spectral lines, and the photoelectric effect), and led to one of the first experimental confirmations of quantum electrodynamics (through the Lamb shift). More recently, several Nobel Prizes in Physics have been awarded for high-precision laser physics, e.g.\ in 1981, 1997, 2005 and 2018.

In parallel, the 20th century also saw the development of a cosmological paradigm, commonly known as the Hot Big Bang model. Its many observational successes are compounded by the fact that about 95 percent of the Universe's contents are in `dark' forms, so far not directly detected in laboratory experiments: our only current evidence for them is indirect, and coming from astrophysical and cosmological observations, subject to various seemingly plausible but possibly unjustified theoretical modeling assumptions. Indeed, the observational evidence for the low redshift acceleration of the universe, accumulated over the last two decades, shows that the canonical cosmological paradigm is, at least, incomplete. It follows that the Hot Big Bang model is at best a simple (though certainly convenient) approximation to the behaviour of a more fundamental cosmological paradigm, yet to be discovered. 

The issue can be summarized by considering the question of whether or not the acceleration of the universe is due to a cosmological constant. This was first introduced by Einstein (in a different historical context) as a mathematical integration constant, but is known to be physically equivalent to a vacuum energy density, as can be obtained in modern quantum field theories. If the acceleration is due to a cosmological constant, the observed value is ca.\ 120 orders of magnitude smaller than the one calculated by particle physicists -- with the obvious implication that such calculations must be incorrect. If it is not, then one needs alternative acceleration mechanisms. Such alternatives exist, the most obvious possibility being cosmological scalar fields, but these will generically violate the Einstein Equivalence Principle. Thus, the evidence for the accelerating universe indicates that there is new physics waiting to be discovered -- either in the quantum field theory sector or the gravitational sector, or possibly in both.

Ongoing developments in the precision, accuracy and stability of astrophysical spectrographs enable them, once again, to play a role in cutting edge fundamental physics, by contributing to the search, identification and characterization of this new physics. This paper discusses the role of ANDES in this quest, focusing on its main science cases (which generally rely on QSO absorption spectroscopy) and emphasizes the requirements for it to play a competitive role in the foreseeable 2030s context (with the U-band coverage being of critical importance), but also discussing synergies with other contemporary facilities. Unless otherwise stated, our analysis relies on version 1.1 (July 2023) of the ANDES Exposure Time Calculator\footnote{Available at \href{https://hires.inaf.it/etc.html}{https://hires.inaf.it/etc.html}}.

ANDES will have the almost unique ability to probe the behaviour of all the fundamental ingredients of the universe: baryons through Big Bang Nucleosynthesis (BBN), radiation (photons) though the temperature of the comic microwave background (CMB), and the dark sector(s), including new dynamical degrees of freedom such as fundamental scalar fields, through tests of the universality of physical laws, probing the stability of nature's fundamental couplings and searching for composition-dependent forces. All these will be complemented by a real-time model-independent mapping of the expansion history of the universe -- a.k.a. the redshift drift. 

Last but not least, one must keep in mind the point that our flagship science cases are photon starved with current facilities, but they will only fully benefit from the ELT's larger collecting area if they are not systematics limited. More specifically, how competitive ANDES will be as a fundamental physics probe will crucially depend on the robustness of its calibration procedures, For this reason, we also briefly discuss the main steps in these procedures and the corresponding requirements.

\section{Big Bang Nucleosynthesis}
% Both Deuterium and He3. 
The light nuclei that were created a few minutes after the Big Bang (a process known as Big Bang Nucleosynthesis -- BBN) currently provide us with the earliest probe to study the physics of the early Universe \citep[for a recent overview, see][]{GrohsFuller2023}. The relative abundances of these nuclei are sensitive to the density of ordinary matter (i.e.\ the baryon density), the expansion rate of the Universe, and the particle content. Therefore, deviations from the Standard Model of particle physics and cosmology can be identified by measuring the relative abundances of the primordial elements, and comparing these abundances to theoretical BBN calculations. There are just five nuclei that are abundantly produced during BBN, including hydrogen (H), deuterium (D), helium-3 ($^{3}$He), helium-4 ($^{4}$He), and lithium-7 ($^{7}$Li). The ANDES design is ideally suited to measuring the primordial abundances of D/H, $^{3}$He/${^4}$He, and $^{7}$Li/H, which all require a high spectral resolution instrument coupled to a large aperture telescope. In order to determine the primordial values of these ratios, we need to discover and observe environments that still retain a primordial composition of the light elements. Furthermore, it is critical to measure the abundances of as many primordial elements as possible; the relative abundances of each element have a different sensitivity to the baryon density, the expansion rate, and the particle content of the Universe. Thus, combining multiple primordial measures allows us to: (1) test the consistency of the cosmological model and the Standard Model of particle physics; and (2) if a departure from the Standard Model is uncovered, multiple primordial abundance measures have the power to unveil the \emph{identity} of the new physics. ANDES will deliver the possibility to revolutionise at least three of the primordial element abundance measures.

\subsection{The primordial deuterium abundance --- D/H}

D/H currently provides our most reliable test of BBN. The standard approach to measure D/H uses near-pristine quasar absorption line systems, typically observed at redshift $z\sim3$, where the lines of interest are shifted into the optical wavelength range (i.e.\ an observed wavelength range of 3600\AA\ -- 5000\AA\, for an absorption line system at $z=3$). In order to accurately model the absorption cloud kinematics, a high resolution spectrograph (such as ANDES) is essential. Based on a sample of just seven systems, the primordial deuterium abundance is known to one percent precision \citep{Cooke2018}, and this sample is not currently limited by systematic uncertainties.

The main challenge of determining the primordial D/H ratio is simply a numbers game. The best systems are: (1) typically those with a higher H\,\textsc{i} column density, log $N$(H\,\textsc{i})/cm$^{-2}\gtrsim19$ (i.e.\ the rarest quasar absorption line systems); and (2) located in the metal-poor tail of the metallicity distribution (making them exceedingly rare!). Combined with this, there are relatively fewer bright quasars than faint ones, so a requisite S/N$\sim20$ is difficult to achieve, particularly at blue optical wavelengths. With the large collecting area of the ELT ($>10\times$ that of the VLT), the community will have access to $\sim100\times$ the number of quasars that are currently possible to observe with the VLT. In concert, this will greatly expand the statistics of deuterium absorption line systems.

\begin{figure*}[h]
\centering
\includegraphics[width=0.32\textwidth]{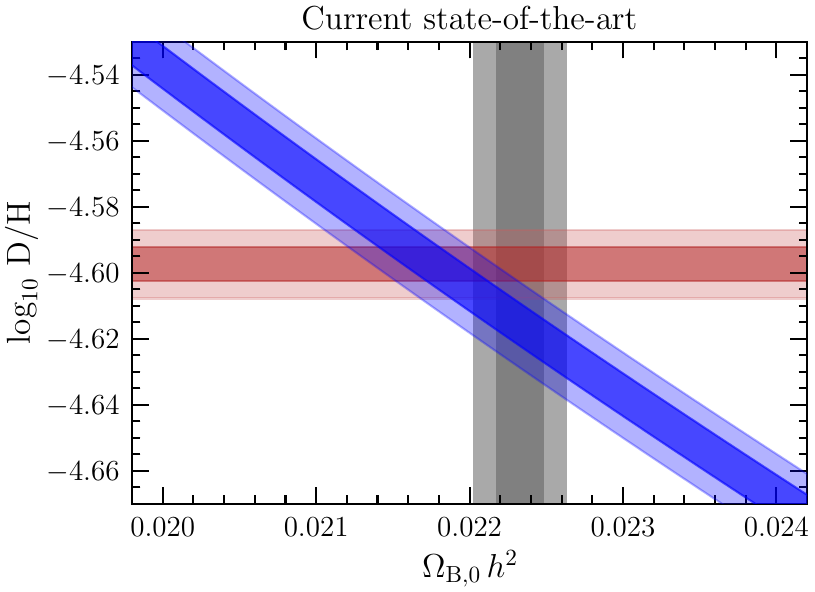}
\includegraphics[width=0.32\textwidth]{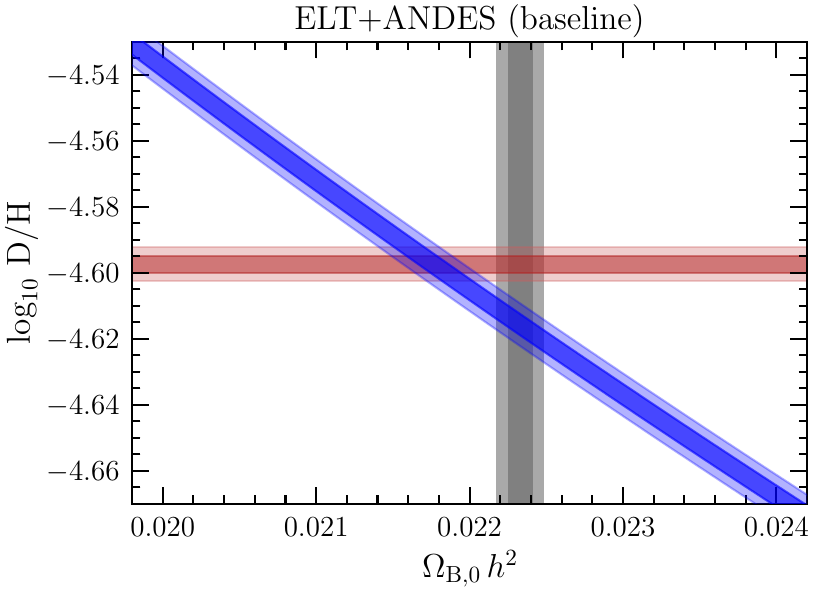}
\includegraphics[width=0.32\textwidth]{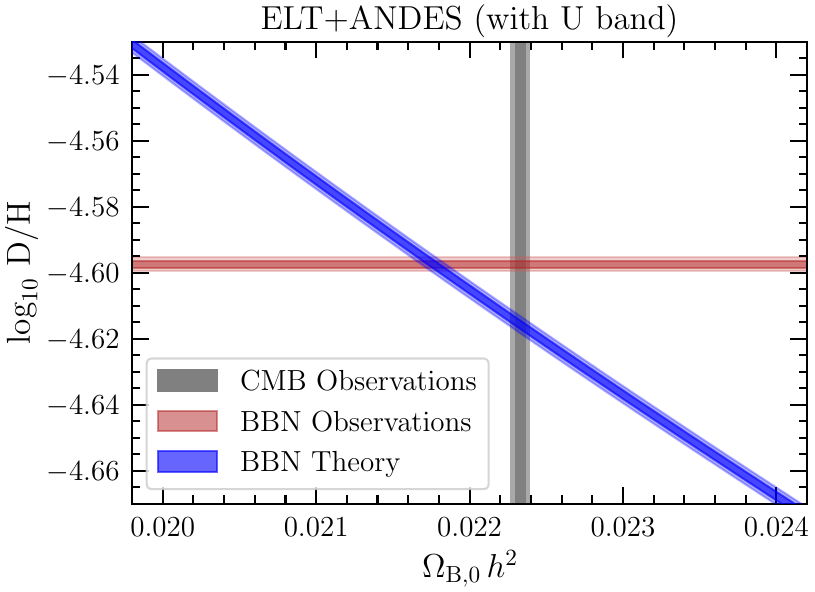}
\caption{The left panel shows the current state-of-the-art quantities that are relevant to BBN ($1\sigma$ and $2\sigma$ confidence regions are shown by the dark and light shades, respectively): (1) the observational determination of D/H (red contours); (2) the baryon density inferred from the CMB temperature fluctuations (grey contours); and (3) the theory conversion between D/H and the baryon density (blue contours). The uncertainties of these three quantities are well-matched with the current state-of-the-art. Assuming that the precision of these measures in the future will reduce by a comparable amount each, the middle panel illustrates the expected constraints from the ELT-ANDES baseline concept (a factor of $\sim2$ improvement of D/H), while the right panel illustrates the benefit of including a U band spectrograph in the ANDES design (a factor of $\sim5$ improvement of the D/H measurement precision).}\label{fig:DH}
\end{figure*}

With the baseline design of ELT-ANDES, \citet{Cooke2016} estimate that the statistics of D/H will be improved by a factor of $\sim2$. These measures will be restricted to somewhat higher redshift D/H absorption systems, and these may suffer from increased blending due to the Ly$\alpha$ forest.
If ELT-ANDES includes a U-band spectrograph (covering 3500--4100\AA),
%; note that a laser frequency comb in the U-band would not be required),
current estimates suggest that the number of D/H measures will be increased by a factor of $>15\times$ \citep[see Figure 7 of][]{Cooke2016}, and make it possible to determine the primordial deuterium abundance to $0.1-0.2$ percent precision (an improvement by a factor of $\sim5\times$ better than current measures). Combining this estimate with theoretical calculations of BBN, this measurement precision corresponds to an uncertainty on the effective number of neutrino species $\Delta N_{\rm eff}=0.02$, which is competitive with CMB stage 4 experiments. Figure~\ref{fig:DH} illustrates the expected improvement to the measurement precision that will be afforded by ELT-ANDES (both the baseline design, as well as the benefit of including a U band extension).

\subsection{The primordial helium isotope ratio --- $^{3}$He/${^4}$He}

Despite being one of the dominant products of BBN, the primordial $^{3}$He abundance has never been determined. Past attempts to determine the primordial $^{3}$He/H ratio primarily used the 8.7 GHz spin-flip transition of $^{3}$He from Galactic H\,\textsc{ii} regions \citep{BalserBania2018}. More recently, it has been appreciated that a reliable determination of primordial helium-3 production can be obtained by measuring the \emph{helium isotope ratio}, $^{3}$He/${^4}$He \citep{Cooke2015}, using absorption line techniques.

In order to measure the helium isotope ratio using He\,\textsc{i}* absorption systems in the local universe, there are three key requirements: (1) coverage of the He\,\textsc{i}* 3889\AA\ and He\,\textsc{i}* 1.083$\mu$m lines; (2) high spectral resolution; and (3) high S/N ratio. The high S/N ratio is required to detect the very weak ${^3}$He\,\textsc{i}* absorption feature. This is particularly needed for the 1.083$\mu$m line, which shows the largest isotope shift ($^{3}$He is shifted by +36.6 km/s relative to ${^4}$He). In order to resolve this weak feature from the neighbouring ${^4}$He\,\textsc{i}* 1.083$\mu$m absorption line (which is stronger by a factor of $\sim 10^{4}$), high spectral resolution is required ($R > 50,000$). Finally, since the corresponding ${^4}$He\,\textsc{i}* 1.083$\mu$m absorption line is very strong, simultaneous coverage of a weaker ${^4}$He\,\textsc{i}* absorption feature is also needed (rest-frame wavelengths of 3889\AA\ and 3188\AA) to reliably pin down the ${^4}$He\,\textsc{i}* column density. The required S/N near 3889\AA\ (S/N$>$50 per 2 km/s pixel) is much less than what is required at 1.083$\mu$m (S/N$>$500 per 2 km/s pixel). We note that this measurement would benefit from the U-band extension,
%(although a laser frequency comb is not required),
but the U-band is not required provided that contemporaneous observations are acquired using a separate U-band spectrograph (e.g.\ VLT-CUBES).

Since the first measurement of $^{3}$He/${^4}$He towards a bright O star in the Orion Nebula \citep{Cooke2022}, more than 40 newly discovered sight-lines have been identified with $^{4}$He\,\textsc{i}* absorption towards stars in the Milky Way and our metal-poor neighbouring galaxy, the Large Magellanic Cloud -- the companion $^{3}$He\,\textsc{i}* absorption has not yet been detected in any of these 40 sight-lines. The main limiting factor of this measurement is that data of especially high signal-to-noise ratio (S/N$>$500 per 2 km/s pixel) are required to detect the weak $^{3}$He\,\textsc{i}* absorption feature. This will be readily achieved with ANDES; for example, the brightest stars in the LMC that are already known to intersect $^{4}$He\,\textsc{i}* absorbing material will reach S/N=1000 in just 30 minutes of ANDES integration (compared to the required $\sim30+$ hours of VLT/CRIRES integration). A fainter (potentially extragalactic) source, such as a cluster of bright stars in a nearby dwarf galaxy, with magnitude $m_{J}$=15 would require just $\sim3$ hours of ANDES integration to reach the requisite S/N=500. Therefore, provided that suitable metal-poor targets can be identified, ANDES will allow the first determination of the primordial $^{3}$He/${^4}$He abundance, and a new test of the Standard Model of particle physics and cosmology.

It may also be possible to detect $^{3}$He\,\textsc{i}* absorption at high redshift; $^{4}$He\,\textsc{i}* absorption has already been identified towards some gamma ray bursts \citep[GRBs;][]{Fynbo2014} and some quasars \citep{Liu2015}. Given the short fading timescales of GRBs and the high S/N required, the large collecting area of the ELT combined with the high spectral resolution afforded by ANDES is the perfect combination.

\subsection{The cosmic lithium problem --- $^{7}$Li/H}

Most determinations of the primordial $^{7}$Li/H abundance are based on observations of the most metal-poor stars in the halo of the Milky Way \citep[e.g.][]{Aguado2019,GonzalezHernandez2019,Mucciarelli2022,Norris2023,Molaro2023}. However, the `primordial' value inferred from these observations is known to be a factor of $\sim3$ lower than the value expected given the cosmological model derived from observations of the CMB. The general consensus is that metal-poor stars burn some fraction of their $^{7}$Li during their lifetime \citep[e.g.][]{Korn2006}, but we cannot yet rule out the possibility that this discrepancy -- also known as the `Cosmic Lithium Problem' -- is a signpost to potentially new physics beyond the Standard Model \citep[e.g.][]{FieldsOlive2022}.

One approach to overcome the issues of stellar burning is to use absorption line observations of gas clouds that sample the chemistry of the interstellar medium \citep{Howk2012}. The larger collecting area provided by the ELT will allow us to probe extragalactic dwarf galaxy environments that are more metal-poor than the SMC. Furthermore, the high spectral resolution of ANDES ($\sim3~{\rm km~s}^{-1}$) will facilitate a greater detection sensitivity, given that the Li\,\textsc{i} absorption lines are usually intrinsically narrow ($\lesssim1~{\rm km~s}^{-1}$).

\section{CMB Temperature}

%Theory
The evolution of the cosmic microwave background temperature ($T_{\rm CMB}$) is a fundamental prediction of the standard model. Any deviation from the expected relation $T_{\rm CMB}(z)=T_{\rm CMB}(0)(1+z)$, which physically relies on the expansion of the Universe being adiabatic and on photon number conservation, would indicate new physics. One example is a violation of the Einstein Equivalence Principle, as a result of a space-time varying fine-structure constant (to be discussed in the next section). Even upper limits to such deviations can be a competitive probe of cosmology, complementing traditional probes such as Type Ia supernovae \citep{Gelo}. 

%Observations
%Local and SZ
Observationally, the CMB temperature has been directly measured at $z=0$ from its black-body spectrum to be $T_{\rm CMB}(0)=2.72548\pm0.00057$\,K \citep{Fixsen2009}. 
At $z>0$, two methods have been developed to indirectly derive this temperature through so-called {\sl thermometers}. The first one relies on inverse Compton scattering of CMB photons by hot intra-cluster gas (Sunyaev-Zeld'ovich (SZ) effect) that induce y-type spectral distortions, whose zero-point depends on the CMB temperature.
This method allows precise measurements \citep[about 7 percent on individual measurements, see][using Planck clusters]{Luzzi2015} but remains limited to $z<1$ where most of the clusters are found. 

%z>1
At higher redshifts, $z\gtrsim1$, the CMB temperature can be measured using absorption lines imprinted by excited fine-structure levels of atomic species or by rotational levels of molecular species in the spectra of background quasars, where the population of the different levels is driven by 
CMB photons\footnote{We note here that, interestingly, \citet{McKellar1941} had already constrained the excitation temperature of CN in our own Galaxy to be as low as 2.3~K, but without giving it much importance as the CMB was only discovered a quarter century later}. The concept is simple, but suitable absorbers are very rare due to the relatively low cross-section of the corresponding cold gas in the ISM of intervening galaxies.
Attempts using electronic (UV) absorption lines from atomic carbon in ground and excited states resulted mostly in upper limits owing to high energy level separation and a dominant contribution to excitation from collisions. 
Observations in the mm-radio domain of the molecular cloud at $z=0.89$ towards PKS\,1830-211 resulted in a measurement with 0.1~K (2 percent) precision, thanks to the observation of various molecular species that allowed tight constraints on physical conditions to be obtained \citep{Muller2013}. 
Finally, electronic absorption lines from CO molecules in different rotational levels provide a unique way to accurately measure $T_{\rm CMB}$ at $z\sim 2-3$ \citep{Noterdaeme2011}. For the majority of the known CO absorption systems, the density remains low enough for collisional excitation to be sub-dominant compared to radiative excitation by CMB photons \citep{Srianand2008, Klimenko2020}. 
While not reaching yet the precision obtained through SZ measurements, these are unique probes of the younger Universe and provide a stronger lever arm for testing the expected linear increase of temperature with redshift\footnote{We note in passing that a measurement has recently been obtained at $z\sim6$ from H$_2$O absorption against the CMB, albeit within a 14\,K-wide 1\,$\sigma$ range \citep{Riechers2022}.}.

%Measuring N(CO,J)  
In practice, several CO bands can be detected at rest-frame wavelengths shorter than 1545{\AA}. The bands are composed of lines from different rotational levels, with Doppler broadening of typically about 1~km\,s$^{-1}$ \citep[e.g.][]{Noterdaeme2017} and a separation between levels $\lesssim$10~km\,s$^{-1}$. In other words, at the $R\sim50,000$ spectral resolution of current observations (VLT/UVES), lines from different rotational levels are partly blended. 
At the twice-higher spectral resolution of ANDES, these lines will be fully separated. This will allow one to obtain reliable constrains on the velocity structure of the gas and to measure the CO column density in each rotational level more independently. 
We note that since this is a line-amplitude measurement, the achieved measurement precision will then depend mostly on the achieved S/N. 

\begin{figure}
    \centering
    \setlength{\tabcolsep}{2pt}
    \begin{tabular}{rl}
        \begin{tabular}{l}
        \includegraphics[width=0.56\hsize, trim=0 20 0 0, clip]{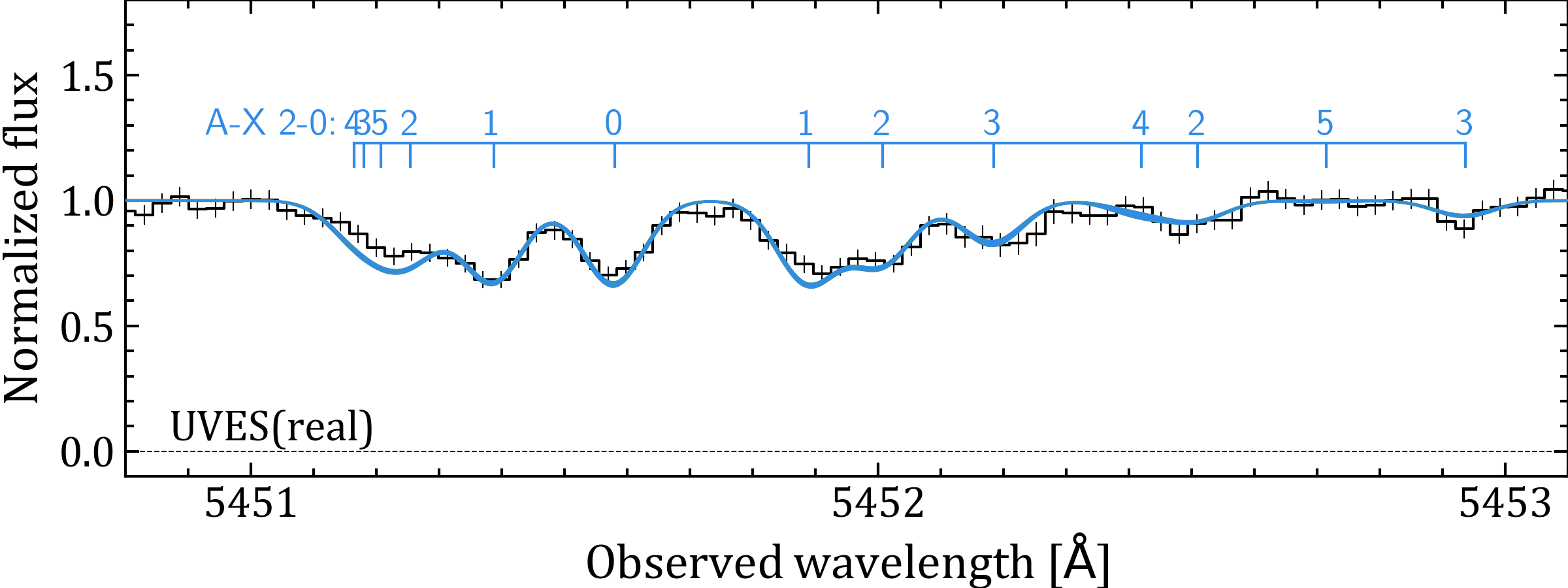} \\ \includegraphics[width=0.56\hsize, trim=0 0 0 0, clip]{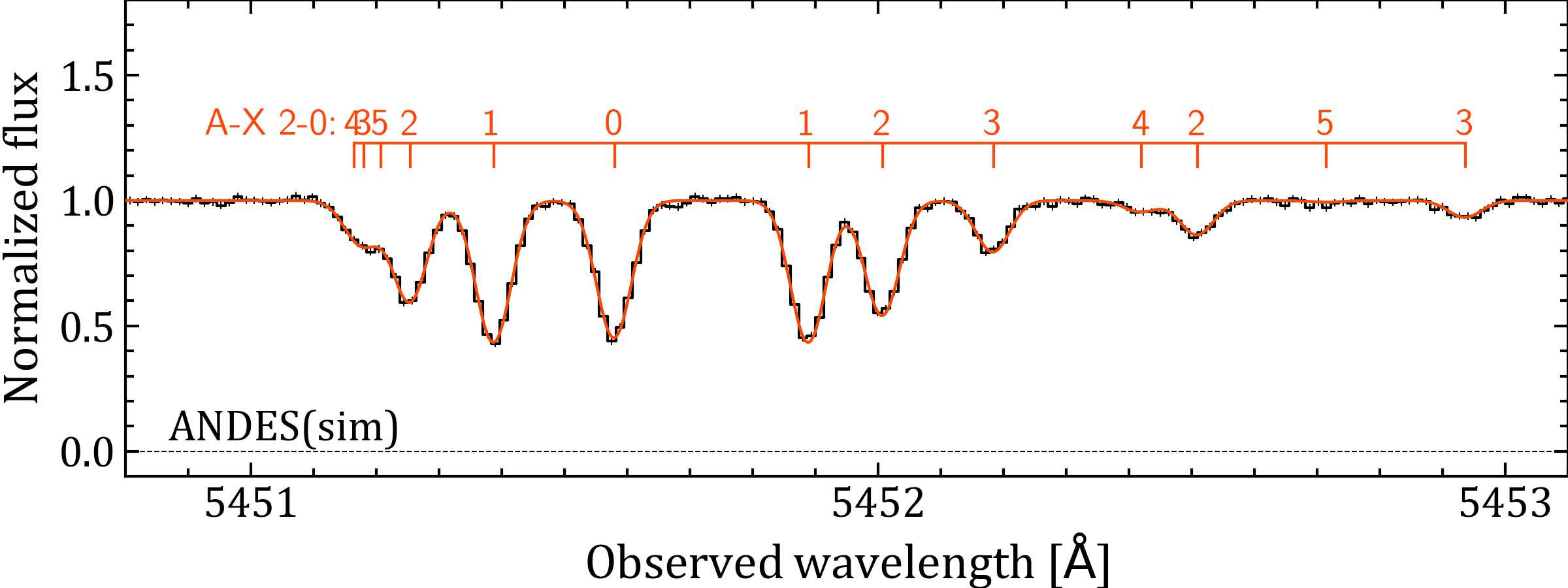}
        \end{tabular}
        & \begin{tabular}{l}
        \includegraphics[width=0.40\hsize]{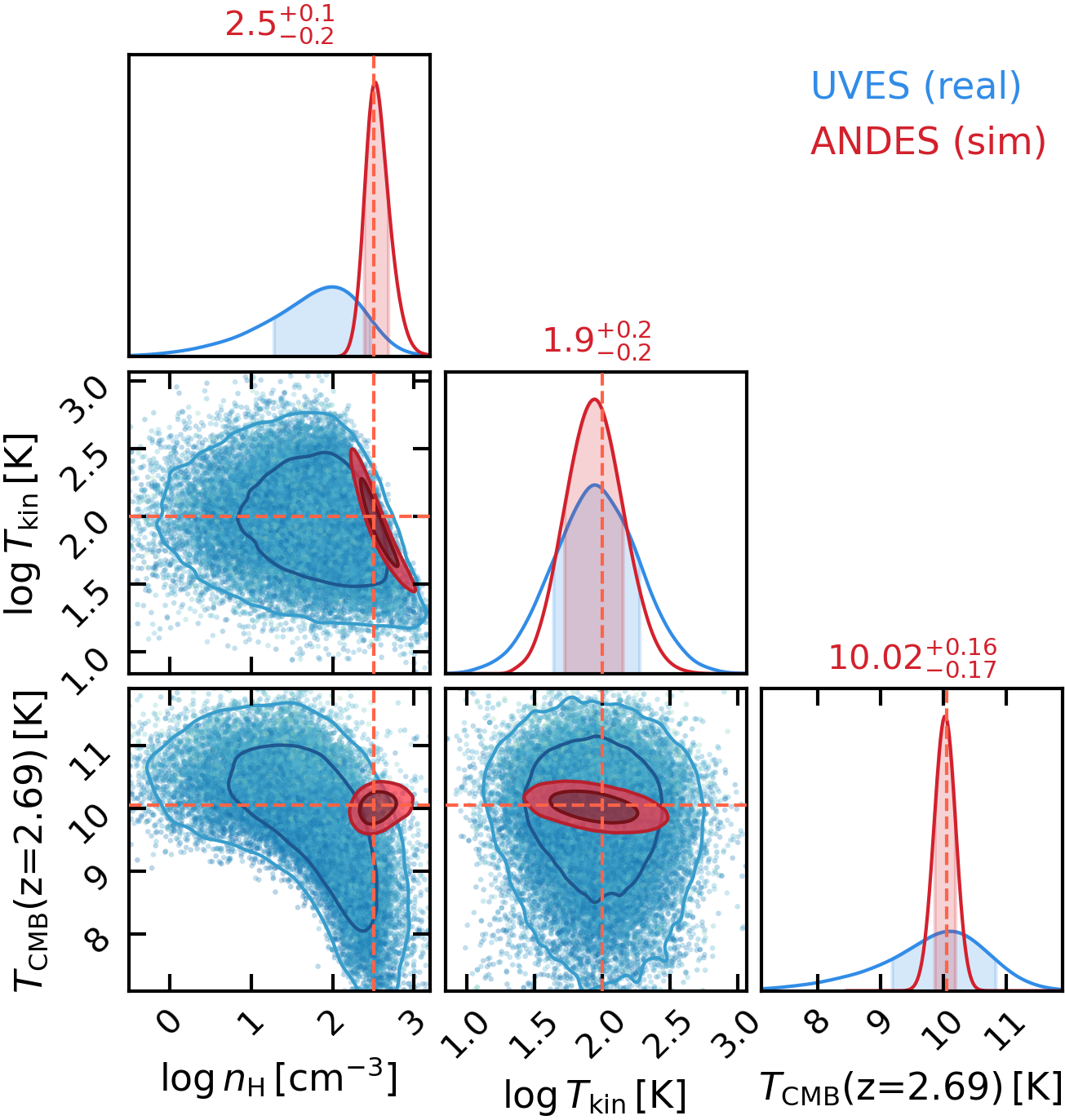}
        \end{tabular}
        \end{tabular}    
    \caption{{\sl Left:} An example of CO absorption band observed at $z=2.69$ towards the quasar J1237+0647 with VLT/UVES ($R=50\,000$, S/N$\sim 40$~pixel$^{-1}$), and as expected with ANDES with $R=100\,000$, S/N$\sim 100$ pixel$^{-1}$. The numbered tick lines indicate the different rotational levels that compose the band. {\sl Right:} Simultaneous constraints on CMB temperature, density and kinetic temperature using profile fitting of CO lines %at $z=2.69$ towards the quasar J1237+0647 
    coupled with excitation model for CO rotational levels. The blue and red colours represent the constraints obtained using existing UVES spectrum and a forecast for ANDES, respectively. %\PN{Replace ESPRESSO's by UVES} \SB{Done, also update the colors}.
    }
    \label{f:COmcmc}
\end{figure}

%ANDES forecast
With ANDES on the ELT, the measurement uncertainties on the CO column densities in each rotational level will become significantly less than the deviation from Boltzmann distribution (i.e.\ radiative excitation alone) due to contribution from collisions \citep{Klimenko2020}. An MCMC analysis of a simulated ANDES spectrum with S/N $\approx$ 100~pixel$^{-1}$ -- which should be reachable in $\sim$30-40~h for an $m=19$ quasar -- 
%{\bf (CHECK...according to ETC, this should be 40h, but scaling from ESPRESSO (x5 S/N, as mentionned in fund. constants section)
%, it should be more like 20h)}-- 
indicates a precision better than 0.2~K at $z=2.7$, i.e.\ smaller than 2 percent uncertainty, see Figure~\ref{f:COmcmc}. Covering the Lyman-Werner bands of H$_2$ and lines from fine-structure levels of C\,{\sc i} from the same target will be very helpful to obtain the relevant priors on the kinetic temperature and number density in the associated medium, respectively. 
We note however that even for the highest-redshift CO absorber known so far \citep[at $z=2.7$ towards J1237+0647, see][]{Noterdaeme2010}, H$_2$ bands will be located in the U-band (3500-4100~{\AA}). Notwithstanding, it remains in principle possible to collect H$_2$ spectra separately using another instrument (e.g.\ CUBES on the VLT)  provided the spectral resolution of the latter allows for precise column density measurements in the low rotational levels.

Finally, CN is in principle an even better thermometer than CO, with its excitation being less sensitive to collisions. This molecule possesses a $BX$ absorption band at 3875~{\AA} rest-frame, which means it could be more comfortably detected using the RIZ ($z<1.45$) or the YJH (at higher $z$) spectrographs. To our knowledge, however, this molecule has not yet been detected through electronic lines at $z>0$.

\section{Varying Fundamental Constants}
\label{Sec:VaryingConstants}

ANDES will be able to compare the values of fundamental physical constants over 12\,Gyr ago and 15\,Gpc away with their current values on Earth. The collecting area of the ELT will provide an unmatched photon-limited uncertainty in these measurements of just 0.3 parts per million (ppm). The challenge is to ensure that instrumental systematic uncertainties can be reduced below this level to provide new, high-precision tests for physics beyond the Standard Model. A detection of variation in any fundamental constant would revolutionise physics, violating the Einstein Equivalence Principle and demand an explanation from a new, more fundamental theory.

\subsection{Motivations and implications for `varying constants'}

Our current understanding of physics is based on a set of physical laws characterised by fundamental constants. However, our best theory of those physical laws -- the Standard Model of Particle Physics -- does not explain the values or origins of these constants. As such, their constancy is a simplifying assumption that must be established or ruled out by experiment. In physically realistic models\footnote{Here we mean models with a low-energy effective action which can plausibly be obtained from a fundamental theory.} containing cosmological scalar fields, they will unavoidably couple to the model's other degrees of freedom, unless a hitherto unknown global symmetry is postulated to suppress the couplings. Thus, one naturally expects such fields to yield varying fundamental couplings and long-range forces \citep{Taylor,Damour1,Damour2}. 

Couplings are experimentally known to {\it run} with energy, and in most standard model extensions they also {\it roll} in time, and {\it ramble} in space -- i.e.\ depend on the local environment, including on other parameters like gravitational potential, dark matter density, dark energy etc.. For example, electromagnetic sector couplings lead to space-time variations of the fine-structure constant, $\alpha \equiv e^2/\hbar c$ \citep{Carroll,Dvali,Chiba}. These theories can guide our intuition about where may be ``better'' to test for variations, but it is often difficult to make reliable predictions across all areas of this broad parameter space and connect constraints from different experiments. It is therefore important to experimentally and/or observationally test for variations in dimensionless fundamental constants in different ways and as a function of as many time-scales, distance-scales and environmental parameters as possible -- see \citep{Uzan,ROPP} for recent reviews of theoretical and observational aspects. Again, null results are highly competitive: current constraints on the variation of $\alpha$ are at the ppm level, while constraints on dynamical dark energy (presumably also due to scalar fields) are at the percent level \citep{Vacher,Schoneberg}. 

Similarly, in beyond-Standard Model theories that unify the four known fundamental interactions, the parameters we call fundamental constants are related to each other, so it is wise to test for variation in more than one constant in these different places, times and environments. Moreover, in most physically realistic models where $\alpha$ varies, the proton-to-electron mass ratio, $\mu\equiv m_p/m_e$, will also vary. The relation between the two variations is model-dependent, but fully calculable in any model that is sufficiently developed to be observationally testable-- see the review by \citet{Uzan} for several examples. ANDES can directly and independently measure $\alpha$ and $\mu$, and therefore has the potential to observationally determine the relation between the two, which would provide a key consistency test of the unification paradigm and potentially rule out a substantial part of the corresponding parameter space.

\subsection{Varying constants in astronomical spectra}

The UV and visible spectra of atoms/ions and molecules are primarily determined by two fundamental constants, $\alpha$ and $\mu$ \citep{Born1935}. The former characterises the coupling strength of electromagnetism, while the latter gauges the chromodynamic scale relative to the electroweak scale \citep[e.g.][]{Flambaum_2004PhRvD..69k5006F}. Importantly, different transitions depend differently on $\alpha$ or $\mu$, so \emph{comparing their relative frequencies in astronomical objects with those found in Earth-based laboratories} is a sensitive test for variations in these constants, even if the astronomical spectra are affected by (much larger) Doppler shifts or cosmological redshifts. For example, the velocity shift of an electronic transition in an atom/ion, relative to its laboratory frequency (wavenumber $\omega$) is \citep[e.g.][]{Dzuba_1999PhRvL..82..888D},
\begin{equation}\label{eq:da}
\frac{\Delta v}{c} \approx -2\frac{q}{\omega}\frac{\Delta\alpha}{\alpha}\,,
\end{equation}
where $\Delta\alpha/\alpha$ ($\ll1$) is the relative difference in the fine-structure constant in the astronomical target and laboratory. Here, $q$ is a sensitivity coefficient which has different magnitudes and signs for different transitions: it depends on the electronic orbital configurations of the transition and must be calculated using advanced numerical quantum mechanics methods \citep[e.g.][]{Dzuba_2002PhRvA..66b2501D,Berengut_2011PhRvA..84e2520B}. A similar relationship characterises the velocity shifts of UV/visible molecular hydrogen and carbon-monoxide transitions \citep[e.g.][]{Ubachs_2016RvMP...88b1003U}.

\begin{figure*}[h]
\centering
\includegraphics[width=0.9\textwidth]{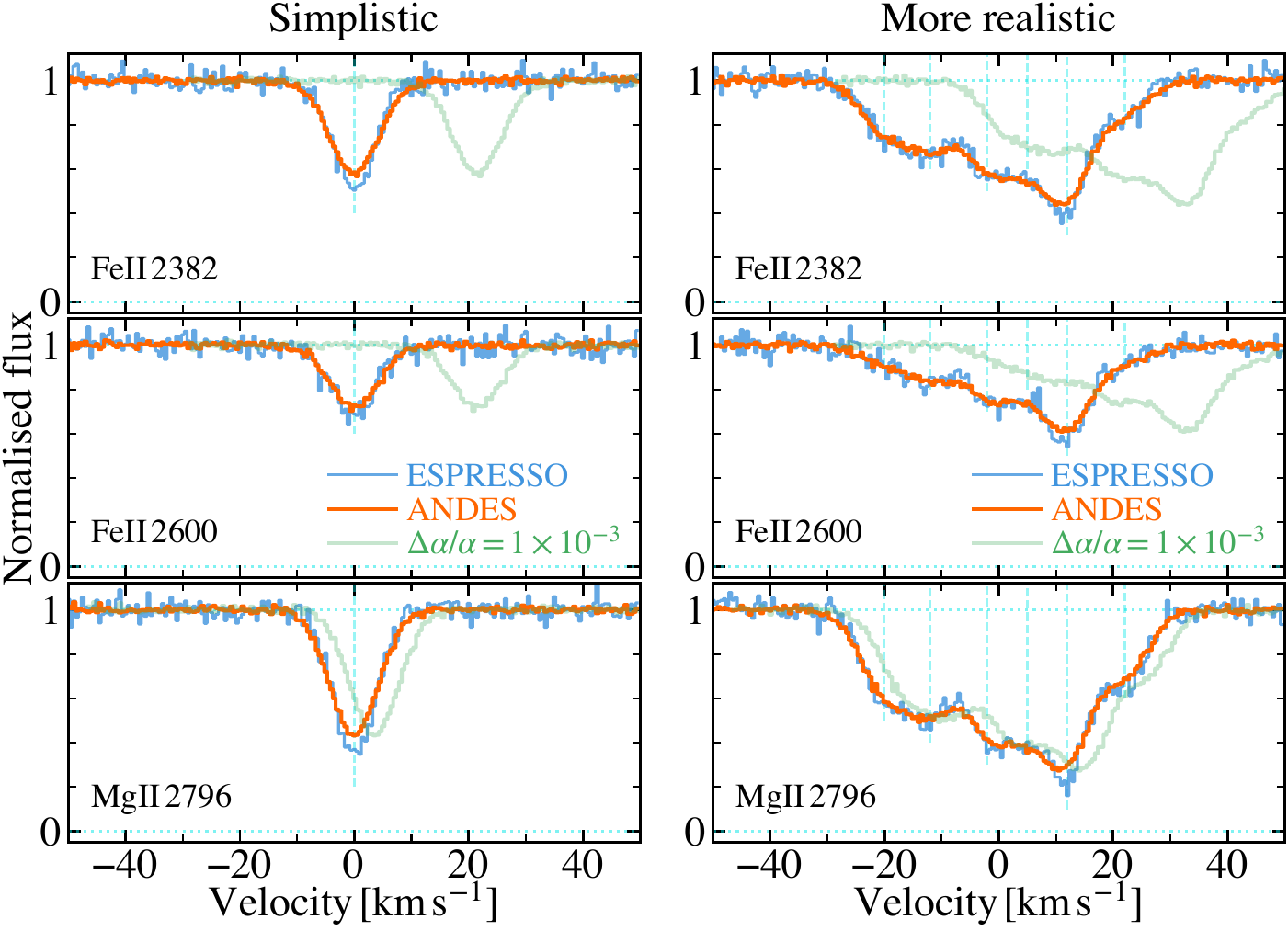}
\caption{Comparison of simulated quasar spectra for ANDES versus ESPRESSO. A 1-hour observation of an $r=17.1$\,mag quasar with ANDES will provide ${\rm S/N}\approx100$ per 3.0\,km\,s$^{-1}$ resolution element, $\approx$3.3 times larger than ESPRESSO. To illustrate the effect of a varying $\alpha$, the simulated ANDES absorption lines (orange) have been shifted by $\Delta\alpha/\alpha=1\times10^{-3}$ to produce the light green lines; this is more than three orders of magnitude larger than the 0.3\,ppm precision on $\Delta\alpha/\alpha$ expected from ANDES. Left panels: An idealised, single velocity component in 3 of the $\sim$5--15 transitions typically observed in quasar absorption systems which can constrain $\Delta\alpha/\alpha$. This would produce a factor of $\sim$5 improvement in statistical uncertainty compared with ESPRESSO. Right panels: A more realistic, multi-component absorber where the higher S/N in ANDES significantly improves the ability to determine the velocity structure as well.}\label{f:simQSO}
\end{figure*}

Figure \ref{f:simQSO} shows how a variation in $\alpha$ would manifest itself in an ANDES spectrum of a quasar absorption system. In the simplistic case of a single velocity component of absorbing gas (left-hand panels), the orange spectrum shows no velocity shifts, i.e.\ $\Delta\alpha/\alpha=0$, between three typical transitions, while the light green spectrum shows those corresponding to $\Delta\alpha/\alpha = +1\times10^{-3}$. This is much larger than the precision achievable with ANDES ($\sim$0.3\,ppm per absorber) but illustrates how transitions of different ions behave differently as $\alpha$ varies. Transitions from the ground states of singly-ionised metals (e.g.\ Mg{\sc \,ii}, Si{\sc \,ii}, Al{\sc \,ii}, Cr{\sc \,ii}, Fe{\sc \,ii}, Zn{\sc \,ii}) at rest-frame wavelengths 1500--2800\,\AA\ are normally the most useful for constraining $\Delta\alpha/\alpha$ in quasar absorbers \citep[e.g.][]{Murphy_2014MNRAS.438..388M}, and typically $\sim$5--15 such transitions are observed per absorber at redshifts $z\sim0.5$--3.5. A similar approach has recently been applied to stellar spectra where, in principle, many more lines could be useful -- see below for further discussion.

Most quasar absorption constraints on variations in $\mu$ have so-far been derived from the Lyman and Werner bands of molecular hydrogen \citep{Ubachs_2016RvMP...88b1003U}. While up to $\sim$100 transitions can be used to constrain $\Delta\mu/\mu$, they fall at rest-frame wavelengths $<$1140\,\AA, necessitating observation at redshifts $z\gtrsim2.9$ where H{\sc \,i} Ly$\alpha$ forest absorption lines blend significantly with the H$_2$ lines, considerably complicating the analysis. The A--X bands of CO (rest-frame $\sim$1300--1550\,\AA) can also be used to constrain $\mu$-variation from optical quasar spectra but, with fewer transitions available than H$_2$ and a similar range in sensitivity coefficients, they typically provide weaker constraints \citep[e.g.][]{Dapra_2017MNRAS.467.3848D}.

\subsection{Observational status of astronomical searches in UV/visible spectra}

Early studies of archival samples of $\sim$50--300 quasar absorbers observed with slit-fed echelle spectrographs (mostly Keck/HIRES and VLT/UVES) showed tentative evidence for deviations in $\alpha$ from the laboratory value at the $\sim$5 parts-per-million (ppm) level, with up to 5$\sigma$ statistical significance \citep[e.g.][]{Webb_1999PhRvL..82..884W,Murphy_2003MNRAS.345..609M}. However, the ThAr lamp calibration of the quasar wavelength scale was \emph{distorted} with respect to that established from the solar spectrum \citep[via reflection from asteroids;][]{Rahmani_2013MNRAS.435..861R}, i.e.\ spurious velocity shifts were applied to different transitions at different wavelengths, most-likely causing the observed deviations in $\alpha$ \citep{Whitmore_2015MNRAS.447..446W}. Subsequent dedicated observations, with corrections for these wavelength distortions from asteroid and solar twin observations, provided tight limits: $\Delta\alpha/\alpha < 1$\,ppm \citep[1$\sigma$; summarised in][]{Murphy_2017MNRAS.471.4930M}. Similar limits were obtained from high-precision studies of a single absorber, towards an exceptionally bright quasar, with fibre-fed laser-frequency comb calibrated spectra from HARPS and ESPRESSO \citep{Milakovic_2021MNRAS.500....1M,Murphy_2022A&A...658A.123M}.

Early H$_2$ constraints on $\Delta\mu/\mu$ from slit-fed spectrographs were also adversely affected by the distortions; once corrected, no evidence for variations in $\mu$ was found from $\sim$10 absorbers at the 2\,ppm level \citep[1$\sigma$;][]{Ubachs_2016RvMP...88b1003U}. No fibre-fed spectrographs have been used for $\mu$ so far, mainly because their efficiency bluewards of $\sim$4000\,\AA\ is very low.

Rather than investigate possible cosmological variations in $\alpha$, recent studies have applied similar techniques to stellar spectra to search for variations within our Galaxy and, especially, near the Galactic Centre which enables a test of any beyond-Standard-Model connection between $\alpha$ and Dark Matter. \citet{Hees_2020PhRvL.124h1101H} compared the velocities of $\sim$10 lines at $\sim$2.2\,$\mu$m in 5 giants, relative to the laboratory values, to limit $\alpha$-variation at the $\sim$6\,ppm level (1$\sigma$). Convective line shifts, which vary from line to line, naturally limit this star-to-lab comparison to that accuracy level. However, by comparing the velocity separations of pairs of lines in the visible \emph{between stars with very similar physical parameters} (and not to the laboratory), these systematic uncertainties are largely removed, providing a limit on variations between stars in the local 50\,pc at the 50 parts per billion (ppb) level \citep{Murphy_2022Sci...378..634M}.

\subsection{Opportunities and challenges for ANDES}

Clearly, ANDES' major advantage will be that the ELT's collecting area will provide a factor of $\approx$3.3 improvement in the \emph{statistical} uncertainty in $\Delta\alpha/\alpha$ available per target, per unit observing time. Crucially though, leveraging that advantage is \emph{only possible if systematic uncertainties remain smaller than the statistical uncertainties from photon noise.} That is, systematic uncertainties from the instrument and calibration must be reduced by a similar factor between ESPRESSO and ANDES.

\begin{figure*}[h]
\centering
\includegraphics[width=0.7\textwidth]{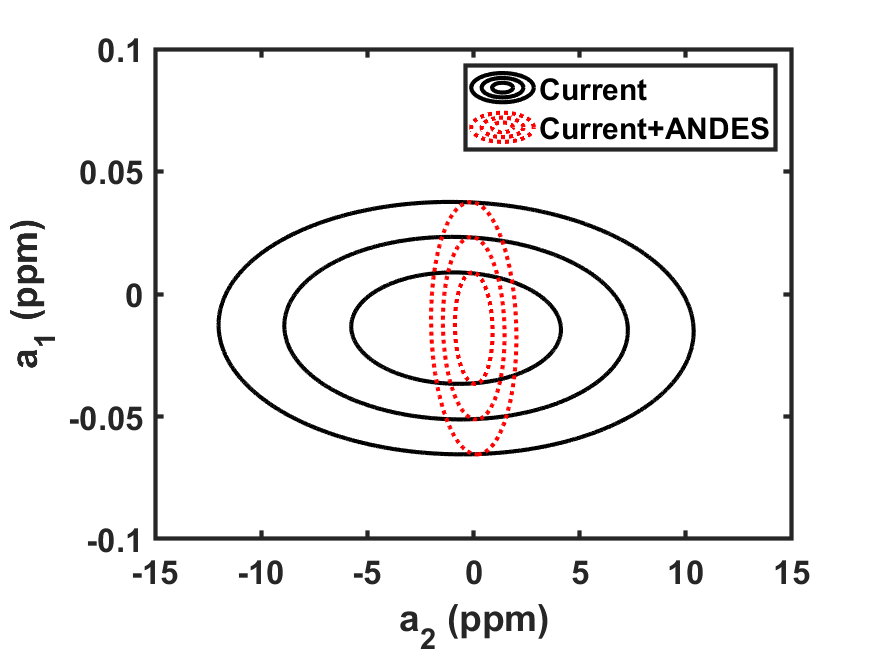}
\caption{Model-independent constraints on the cosmographic series for $\Delta\alpha/\alpha$, cf.\ \citet{AlphaCosmography}. The coefficient $a_1$ is predominantly constrained by local experiments (e.g., atomic clock tests), while $a_2$ can only be constrained by astrophysical observations. Solid black contours show current constraints, and red dotted ones show the impact of ANDES measurements of 15 absorbers, under the assumptions outlined in the text, together with the assumption of a fiducial model without variations.}\label{f:cosmography}
\end{figure*}

For quasar absorption constraints, the most important systematics are those that spuriously shift one transition's measured velocity relative to another. The $\sim$1\,ppm statistical uncertainties produced from single absorbers, or small samples, with 8-to-10-m class telescopes imply that ANDES will achieve $\sim$0.3\,ppm uncertainties, corresponding to $\sim$6\,m\,s$^{-1}$ relative line shifts. That is, in a single quasar exposure, the wavelength scale and all instrumental effects that affect the measured centroid of an absorption line (e.g.\ instrumental profile variations, charge transfer inefficiencies, etc.) must allow the relative velocities of lines at different wavelengths to be measured with $\sim$1\,m\,s$^{-1}$ \emph{accuracy} (regardless of the photon statistical noise in the lines). Even with laser frequency comb calibration, ESPRESSO's accuracy in this context is $\sim$20\,m\,s$^{-1}$ \citep{Schmidt_2021A&A...646A.144S}, so improvements will be required. It should also be mentioned that the (often complex) velocity structure and unknown relative isotopic abundances of metal-line absorbers also present challenges and potential systematic errors as well, with ongoing investigations into their effect on $\Delta\alpha/\alpha$ \citep[e.g.][]{Murphy_2004LNP...648..131M,Lee_2021MNRAS.507...27L}. Assuming that ANDES is not limited by systematics, Figure~\ref{f:cosmography} illustrates, in a model-independent way, the potential cosmological impact of its $\alpha$ measurements.

To enable new constraints on $\mu$-variation with $\sim$0.5\,ppm uncertainties, the proposed $U$-band channel of ANDES will be crucial. Its 3500--4100\,\AA\ wavelength range is required to cover an adequate number of H$_2$ lines for absorbers at $z\lesssim3.1$, i.e.\ 9 of the 10 known systems for which precise measurements are possible. However, a somewhat less stringent relative velocity systematic error budget of $\sim$5\,m\,s$^{-1}$ is allowable across the $U$-band. The ANDES baseline design could constrain $\Delta\mu/\mu$ with $\sim$1\,ppm precision in the 10th known system (at $z=4.22$). Identifying new H$_2$ absorbers at $z\gtrsim3.1$ would be very important for improving the constraints below this uncertainty level, but the faintness and lower number-density of quasars, as well as increased blending due to the Ly$\alpha$ forest, at these higher redshifts all work against achieving substantial improvements.

The ability to measure $\mu$ in multiple systems at a broad range of redshifts will endow ANDES with another unique scientific opportunity. When measuring $\mu$ using H$_2$ one is indeed measuring $m_p/m_e$; however, when using other molecules, such as CO, whose nuclei contain neutrons, one is measuring the ratio of an effective nucleon mass to the electron mass, which will coincide with $m_p/m_e$ in the standard model, but not in any extension of the standard model where there are composition-dependent forces (in other words, scalar fields with different couplings to protons and neutrons). Thus measurements of $\mu$ using two or more molecules in the same absorption cloud are an astrophysical test of composition-dependent forces\footnote{Conceptually, this is akin to checking whether one proton and one neutron fall at the same rate under gravity.}.

Given that the differential stellar measurements of $\Delta\alpha/\alpha$ are very recent and, it seems, currently not limited by instrumental systematics \citep{Berke_2023MNRAS.519.1221B}, it is less clear whether improving upon them will imply specific requirements upon ANDES. However, that method is currently limited to \emph{close pairs} of lines only because of the potential for wavelength calibration and instrumental errors when comparing lines in very different parts of the detector plane. To overcome that limitation, one must be able to compare the velocities of lines with large separations \emph{between stars}. It is not necessary that these relative velocities are correct, but only that they remain the same between exposures of different stars one wishes to compare. The requirements for the quasar absorption-line constraints on $\alpha$ above should be sufficient for this provided they are maintained over reasonable periods of time (e.g.\ years).

\section{Redshift Drift}
\label{Sec:RedshiftDrift}

The discovery that the expansion of the Universe began accelerating $\sim$5\,Gyr ago \citep{Riess98,Perlmutter99} was awarded the Nobel prize in physics in 2011 because it clearly indicates the existence of new physics beyond the Standard Model. While still lacking an accepted physical explanation, the acceleration is usually phenomenologically modelled by introducing a new energy component, called `dark energy', with the unusual property of having an equation of state parameter $w_{\rm DE} = p_{\rm DE} / (c^2 \rho_{\rm DE}) < -1/3$. Over the past two decades, a large amount of observational effort has been, and is continued to be, expended on determining the value of $w_{\rm DE}$ and whether it varies with redshift, using a variety of cosmological probes. So far, all observations appear to be consistent with the acceleration being caused by the simplest possible form of dark energy, i.e., a cosmological constant $\Lambda$ (corresponding to $w_{\rm DE} = -1$) \citep[e.g.][]{Escamilla23}, comprising of about $70$ percent of the Universe's total energy, thus leading to the now standard $\Lambda$CDM model of cosmology.

With the accelerated expansion still remaining as the only observational/experimental evidence of dark energy so far (a state of affairs that may be changed by ANDES, as explained above), measuring the expansion history with as much precision and with as many different kinds of observations as possible continues to be a key objective of present-day cosmology. \citet{Sandage62} first pointed out that the evolution of the expansion rate causes the redshift of a cosmologically distant object to slowly drift in time: $\dot z(z) = (1+z)H_0 - H(z)$, where $H$ is the Hubble-Lema\^itre parameter. This implies, at least conceptually, that a simple spectroscopic monitoring campaign over some time-scale $\Delta t$ of a number of objects distributed over a range of redshifts could reveal the expansion history. The difficulty with this approach is of course the smallness of the effect: $\Delta z = \dot z \Delta t \approx H_0 \Delta t \approx 10^{-9}$ or $\sim$$6$\,cm/s over a decade.

Despite this dauntingly small number, there are good reasons to nevertheless pursue a measurement of this effect. Since the derivation of the above relation between the redshift drift and the expansion history $H(z)$ only relies on the cosmological principle and on the assumption that gravity can be described by a metric theory, the redshift drift offers an entirely direct and model-independent route to the expansion history. Uniquely among all cosmological experiments, the redshift drift infers the expansion history by a comparison of two different past light cones and thus represents an entirely non-geometric probe of the global dynamics of the metric. Furthermore, it does not involve any assumptions about the astrophysics of the sources involved.

\citet{Liske08} first demonstrated that a measurement of the redshift drift was in principle within reach of the ELT. Assuming a purely photon-noise limited experiment, i.e.\ excluding all instrumental systematics, they showed that the effect could be detected by monitoring the redshifts of Ly$\alpha$ forest and other absorption lines in the spectra of the $\sim$10 brightest known quasars in the redshift range $2 \lesssim z \lesssim 5$ over a period of $\sim$20~years. What exactly can be achieved by such observations depends sensitively not only on the brightness and redshifts of the available quasars, the efficiencies of the ELT and ANDES, and the amount of observing time one is willing to spend, but also on the exact selection of the quasars used for the experiment \citep{Esteves21} and any external priors (e.g.\ on $H_0$ and/or flatness) one brings to bear. Recently, the QUBRICS survey discovered a set of previously unknown bright quasars in the southern hemisphere \citep{Cristiani23}, improving our target selection options, while \citet{Cooke20} suggested the additional measurement of the differential redshift drift using the `Ly$\alpha$ cell' technique as a methodological improvement. However, despite these advances, and considering a realistic ANDES efficiency, the detection of the redshift drift will likely cost a few 1000 hours of observing time spread over 20--25 years.\footnote{A potential alternative version of the redshift drift experiment that does not require a $\sim$two-decade experiment duration consists of measuring the redshift difference between the multiple images of a gravitationally lensed source \citep{Wang23}.}

Although a redshift drift measurement from ANDES by itself is thus unlikely to yield precision constraints on cosmological parameters, its combination with similar measurements at $z < 2$ expected from SKA \citep{Kloeckner15} and CHIME \citep{Yu14} will be extremely valuable. The role of ANDES in this partnership is to fix the expansion history in the matter-dominated era in order to allow the low-redshift measurements to constrain $w_{\rm DE}$ \citep{Alves19,Rocha23}. Moreover, the redshift drift constraints in the $w_0$-$w_a$ plane of the CPL parametrization tend to lie perpendicular to those of more canonical cosmological probes, thus facilitating the breaking of their degeneracy \citep{Martins21}. 

In summary, ANDES will be able to provide us, for the very first time, with an entirely new and unique route to the expansion history of the Universe at $z > 2$. The model-independent nature of a redshift drift measurement will probe the cosmological paradigm at a very fundamental level and, in synergy with traditional cosmological probes, redshift drift measurements from other facilities, and the other fundamental physics experiments from ANDES itself described above, will help to shed light on the nature of dark energy. 

We emphasize that this science case is not only extremely photon-starved but also places rather demanding requirements on the wavelength calibration of ANDES, in particular on its stability (as defined in Figure~\ref{Fig:Visualization_PrecisionAccuracyStability} below). Fully capitalising on the ELT's photon collecting power and enabling a total (i.e., summed over all observations) radial velocity precision at the cm/s level, obviously requires that this precision improves with the number of observations as $1/\sqrt{N}$ right down to this level. Having to collect these observations over a time-scale of $\sim$two decades, implies that the bias of the wavelength solution that remains after all calibrations have been applied, may not drift by more than $\sim$1~cm/s over this time-scale (see Figure~\ref{Fig:Visualization_PrecisionAccuracyStability}). As further discussed below, this is a challenging specification -- one that requires substantial research before it can be met. For this reason, the redshift drift science case and its associated demand on the stability of ANDES are only considered goals (as opposed to strict requirements) during the design of ANDES.

Finally, we note that the redshift drift is, at least to some extent, a representative and gateway for other science cases addressing the weak gravity regime. Comparing velocity measurements with $\sim$cm/s precision over the time-scale of a decade corresponds to an acceleration measurement with a precision of $\sim$$10^{-10}\,$m/s$^2$, which is precisely the scale at which dark matter is required to explain the observed dynamics of galaxies and clusters of galaxies. ANDES will thus enable the ELT to not only probe gravity in the strong-field regime in the vicinity of the supermassive black hole in the centre of the Milky Way using MICADO \citep{Davies21}, but also in the extreme weak-field regime.

\section{Required research to improve the wavelength calibration and to enable key science cases}
\label{Sec:WaveCal}

\begin{figure*}[h]
 \includegraphics[width=\linewidth]{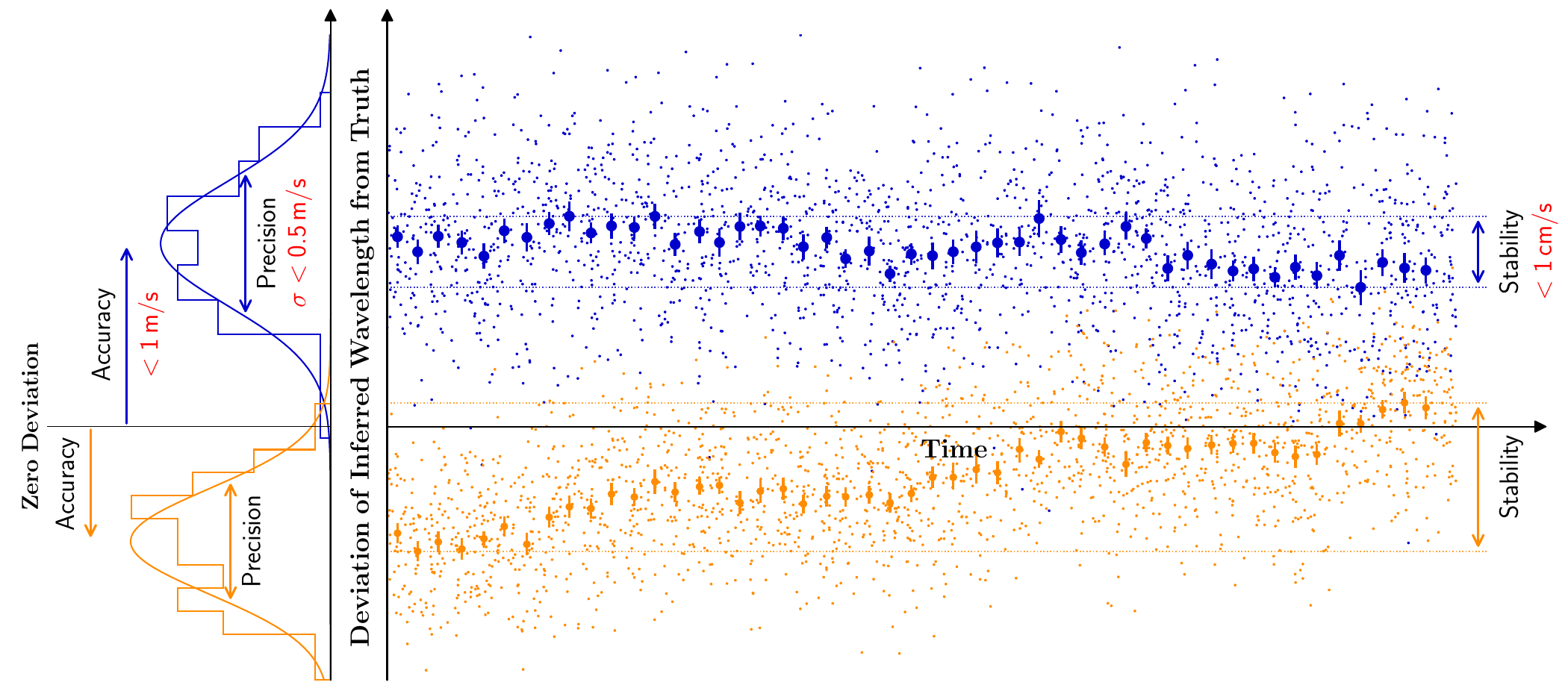}
 \caption{
 Visualization of the adopted concepts for \textit{precision}, \textit{accuracy}, and \textit{stability}.
 The right panel shows a sequence of wavelength measurements (small dots) for two spectral lines (blue and orange) taken over a certain period of time. Shown is the deviation of the inferred value from the truth (horizontal black line). To aid the eye, measurements are grouped into bins (thick dots) and the left panel shows histograms for the points falling into the first bin.
 \textit{Precision} is the scatter of formally identical measurements (due to stochastic effects), while \textit{accuracy} describes the offset (or bias) of this distribution from the true value.
 Over time, this offset will not be stable but be subject to drifts, which characterizes the \textit{stability}.
 The ANDES design goals for these properties are indicated in red. 
 }
 \label{Fig:Visualization_PrecisionAccuracyStability}
\end{figure*}

The requirements for wavelength calibration imposed in particular by the science cases related to \textit{Varying Constants} (Section~\ref{Sec:VaryingConstants}) and \textit{Redshift Drift} (Section~\ref{Sec:RedshiftDrift}) are so demanding that precise and accurate wavelength calibration of the spectrograph becomes a research topic of its own. As a direct consequence of the larger collecting area of the ELT, ANDES needs to be calibrated nearly one order of magnitude better than the best extreme-precision RV spectrographs on 8\,--\,10\,m class telescopes, including the current flagship for stable and accurate measurements, VLT/ESPRESSO. Only with improved wavelength calibration will ANDES be able to benefit from its huge photon-gathering capacity and avoid being limited by systematics. Achieving this leap forward in wavelength calibration accuracy and stability (Figure~\ref{Fig:Visualization_PrecisionAccuracyStability}) requires active and dedicated research on this topic.
Here, ESPRESSO acts in many ways as a benchmark and testbed for the development of new calibration methods and techniques.

\subsection{Classic ThAr/FP wavelength calibration}

Wavelength calibration of high-resolution echelle spectrographs is usually based on hollow cathode lamps  (HCLs), either ThAr or UNe, augmented by passively-stabilized white-light Fabry-P\'erot etalons (FPs). Combining information from both of these highly complementary sources provides a high-quality wavelength solution \citep[e.g.][]{Cersullo2019} and will also be the baseline for ANDES. Although probably not sufficient to achieve all wavelength calibration requirements, it is worth improving this very reliable and cost-efficient calibration method.

Recently, significant advancements in the understanding of FPs have been achieved, in particular related to their \textit{chromatic drift} \citep{Terrien2021, Schmidt2022, Kreider2022}. By correcting for this effect, a much improved wavelength calibration could be achieved, and the new ESPRESSO DRS 3.0.0 now shows outstanding performance.
Still, the fundamental limitation for the calibration via HCLs and passively stabilized FPs is that for long timescales (longer than a few days) all \textit{absolute} wavelength information is provided by the ThAr or UNe spectra. These, however, have limited information content due to the sparse and unevenly bright lines, suffer from blending of lines, and do not provide truly fundamental wavelength calibration but rely on laboratory measurements, which currently seem not accurate enough to achieve the $1\,\mps$ accuracy goal. In addition, HCLs are subject to wear, which limits their lifetime and causes a drift of the lines \citep{Dumusque2021}, compromising the extremely challenging goal of $1~\cmps$ stability over decades. Thus, a better characterization of the HCL behavior could substantially improve the accuracy and stability of the classic ThAr/FP calibration.

\subsection{Laser frequency combs}

A totally different type of calibration source that can in principle overcome these limitations are laser frequency combs \citep[LFCs;][]{Murphy_2007MNRAS.380..839M}. These devices produce a dense ensemble of lines with accurately known frequencies, locked against a reference from an atomic clock. % \citep{Reichert1999, Jones2000, Udem2002}.
Substantial efforts have been put into the development and demonstration of LFCs at astronomical spectrographs 
%\citep[e.g][]{Wilken2010a, Wilken2012, Phillips2012a, Ycas2012, Chang2014, Probst2014, Depagne2016, McCracken2017a, Obrzud2018, Ludwig2023}.
\citep[see e.g][]{Fischer2016}.
However, LFCs are extremely complex devices and come with many difficulties, in particular when adopting them to the needs of astronomical spectrograph calibration, e.g.\ in terms of line separation and wavelength coverage (at which point these special LFCs are often referred to as \textit{astrocombs}).
Therefore, only few observatories currently use them for routine calibration \citep[e.g.][]{Kokubo2016, Metcalf2019b, Hirano2020, Terrien2021, Blackman2020}. In most cases, they remain somewhat experimental and in particular the LFCs at HARPS and ESPRESSO have experience several technical problems. Substantial improvements have to be made to bring astrocombs to the technical readiness level required for routine and reliable science operations with ANDES.
In addition, novel technologies are needed to provide LFC coverage for the full ANDES wavelength range, in particular the short-wavelength part for which no astrocombs exist so far (\textit{U} and \textit{B} band). Developments in this direction are ongoing \citep[e.g.][]{Wu2023, Ludwig2023, Cheng2023}, but these technology demonstrations are still quite far from a system suitable for routine operation.

\subsection{Data reduction}

In addition, new methods for data reduction, calibration, and analysis are needed. In particular for the wavelength calibration accuracy (needed for the determination of the fine-structure constant), a very careful modeling of the instrumental line-spread function (LSF) is required. Studies with ESPRESSO have revealed that intra-order distortions of the wavelength solution by up to 50\,m/s can be introduced when not properly accounting for the highly non-Gaussian LSF \citep{Schmidt_2021A&A...646A.144S}. Preliminary results show that a careful measurement and modeling of the ESPRESSO LSF can reduce the intra-order distortions and internal inconsistencies to the few m/s level. This is a major step forward, but still not quite sufficient to meet the ANDES requirements.
One of the limitations identified so far lies in the measurement of the LSF.
This requires a calibration source with truly unresolved lines. Currently, the ESPRESSO LFC is used for this, but its lines have fixed frequencies and therefore always fall on the same spots on the detector, providing only a very pixelated measurement of the LSF. To overcome this sampling issue, tunability of the LFC is highly desired and, in principle, advertised by manufacturers but so far not possible. As an alternative, it is currently being investigated whether a dedicated high-finesse Fabry-P\'erot etalon could be a viable and simpler solution to facilitate accurate LSF determination.

Recently, it was also discovered that even for wavelength calibration stability (ignoring accuracy aspects), an accurate modeling of the LSF is inevitable. Tests with ESPRESSO revealed apparent shifts of the LFC lines up to $\pm2\,\mps$ associated with variation in the LFC flux. With a proper forward-modeling of the LSF, this issue could be substantially reduced, but still, the calibration derived from the LFC calibrations were substantially less stable than the one from the FP. Due to these issues, variations of the LFC flux and limitations in the data processing, it is thus currently not possible to fully benefit from the LFC calibrations. Improvements in both aspects are needed to unleash the full potential of the LFC calibration.
On a similar note, \citet{Probst2016} and \citet{Milakovic2020a} reported a systematic inconsistency of $\approx50\,\cmps$ between two similar LFCs that were at some point simultaneously installed at HARPS. This is another clear indication that without more advanced data processing tools the quality of the derived wavelength calibration falls significantly short of the theoretical performance of the LFC and that more work is needed to adapt the methods to the special properties of LFCs and use their full potential.

This even raises the question whether classical spectral extraction algorithms like~\citet{Zechmeister2014}, used for ESPRESSO, or~\citet{Piskunov2021}, which is the current baseline for ANDES, are actually capable of delivering spectra with the required accuracy and stability. Like most other approaches, these algorithms make the fundamental assumption that the two-dimensional instrumental profile (IP) can be decomposed into two one-dimensional functions, which massively simplifies the extraction process, but since never fully satisfied, inevitably introduces residuals and systematics. Their amplitude depends on the actual IP shape and how close the assumed shape is to the truth.
It might become necessary to implement \textit{spectro-perfectionism} \citep{Bolton2010} for ANDES, an algorithm that models the two-dimensional flux distribution on the detector as superposition of individual IPs. Although extremely powerful, this approach imposes strong demands not only on the computational power needed for data reduction but also in terms of specialized calibration light sources required to accurately characterize the IP in the first place.

\subsection{Validation of the wavelength calibration}

Another major challenge is the validation of the wavelength calibration to later convince the scientific community that the measured effects (change in $\alpha$ or redshift drift) are genuine and not instrument-induced. In particular for the redshift drift and the fundamental constants experiment, there will be no independent verification on sky, since there is only one Universe to observe, and the data itself will not inherently tell whether the measurement is true or false. Furthermore, the search for varying constants might in general remain a null-measurement with the goal to provide the tightest constrains possible, an inherently extremely difficult task. Therefore, it is crucial to also provide validation tests, in addition to the science observations.

Significant efforts are therefore underway to develop techniques for independent validation of the wavelength calibration. One of the concerns here is that the calibration unit injects calibration light at the spectrograph front end. Light from science targets, however, also passes through atmosphere and telescope and is therefore unavoidably injected into the optical fibers in a slightly different manner. The optical design of the instrument contains elements that reduce the susceptibility to the fiber injection as much as possible, e.g.\ by the use of \textit{double scramblers} %\citep{Hunter1992} 
in the optical fiber train, but given the extreme requirements (accuracy, stability, timescale), even very minute effects could compromise the measurements and would not be calibratable. Therefore, suitable wavelength validation sources are needed that are located upstream of the telescope, ideally on sky. Naturally occurring astrophysical sources can not be used since none provides sufficient wavelength accuracy or stability.

One solution could be the use of gas absorption cells, in particular I$_2$ cells. In contrast to e.g.\ \citet{Butler1996}, these shall not be placed into the beam when observing science targets (this would be extremely inefficient), but instead only used to verify the wavelength solution of the spectrograph, derived from internal calibration sources, against the I$_2$ spectrum. For this, iodine cell observations of bright, featureless stars are sufficient, creating an artificial star with well controlled (and independently measurable) absorption features. This probes the full optical path from the top of the atmosphere to the detectors and is therefore equivalent to actual science observations. The spectrum of the iodine cell itself is expected to be extremely stable and can be accurately measured in the laboratory.
Recently, a proof-of-concept experiment has been conducted at ESPRESSO to explore the feasibility and achievable precision of this approach.
After the data analysis is completed, a more detailed long-term monitoring with ESPRESSO is planned and, if proven viable, the same strategy will be implemented for ANDES.

Other, but more exotic, concepts to validate the wavelength calibration include observations of the Raman spectrum of the adaptive optic facility laser-guide stars \citep{Vogt2019,Vogt2023}, calibration sources carried by drones, or even laser frequency combs installed on satellites. Significant research and development over the next years is necessary to demonstrate the practical application of these concepts.

\subsection{Further steps}

Over the last few years, much valuable experience has been gained with ESPRESSO which has revealed numerous systematics that, at a certain point, may become limiting factors. To meet the much more demanding ANDES wavelength calibration requirements, substantial advancements have to be made and these systematics characterized, understood and eliminated, either by an improved design of the instrument or -- more likely -- by advanced data processing and calibration procedures.
There are still several years of time for the development of advanced data processing algorithms, however, all aspects related to the hardware design have to be addressed right now during the design phase of the instrument. One of the most important aspects here is to define which type of calibration sources are needed and how they shall be operated to later facilitate a precise, accurate, and stable wavelength calibration of ANDES. Therefore, various efforts are ongoing to gain further experience in advanced wavelength calibration techniques, to push the boundaries towards the ANDES goals, and to incorporate this knowledge into the ANDES design.

\section{Outlook}

We have discussed the role of ANDES in cosmology and fundamental physics, specifically in probing the physical behaviour of all the fundamental ingredients of the universe -- baryons, radiation, and the dark sector(s), which presumably include new dynamical degrees of freedom beyond the standard cosmological model -- as well as in carrying out several hitherto unexplored consistency tests. While the outcome of this exploration is not known a priori, we can nevertheless outline two main scenarios.

The first scenario is the null case, in which no deviations from the behaviour expected in the standard $\Lambda$CDM model are found, but one simply improves currently available limits. It must be emphasized that this \textit{minimum guaranteed science} can be highly competitive with more traditional cosmological probes. As an example, constraints on scalar field dynamics derived from varying $\alpha$ measurements are currently at the ppm level, while those derived from cosmological data on the dark energy equation of state are at the percent level \citep{Vacher,Schoneberg}. Moreover, this minimum guaranteed science is calculable a priori, in an approximately model-independent way, simply from the ELT and ANDES technical specifications, at least if one can assume that the observations are not limited by systematics -- in other words, if ANDES can make full use of the photons which the ELT provides.

The second scenario ensues if one or more such deviations are detected, e.g.\ mutually incompatible BBN abundances, a violation of the temperature-redshift relation or of the Einstein Equivalence Principle, or possibly a value of the redshift drift incompatible with the $\Lambda$CDM prediction for the corresponding redshift. While the impact of such a detection would be obvious \citep{Taylor,Damour1,Damour2}, it is important to note that it will also significantly impact ANDES' own observing strategy. If one is aiming to obtain the best possible limits, a simple and reasonable strategy (though one with some model dependence) is to concentrate the available telescope time on a small number of bright targets. On the other hand, if, for example, a varying $\alpha$ is detected, one would obviously like to map the redshift dependence of this variation using the widest possible range of redshifts, from $z<1$ (when the putative dark energy is dynamically relevant) to deep in the matter era -- a fine-structure survey, as recently suggested by \citet{Peebles}. A more detailed exploration of such observational strategies is left for subsequent work.

\begin{figure*}[h]
\centering
\includegraphics[width=1.0\textwidth]{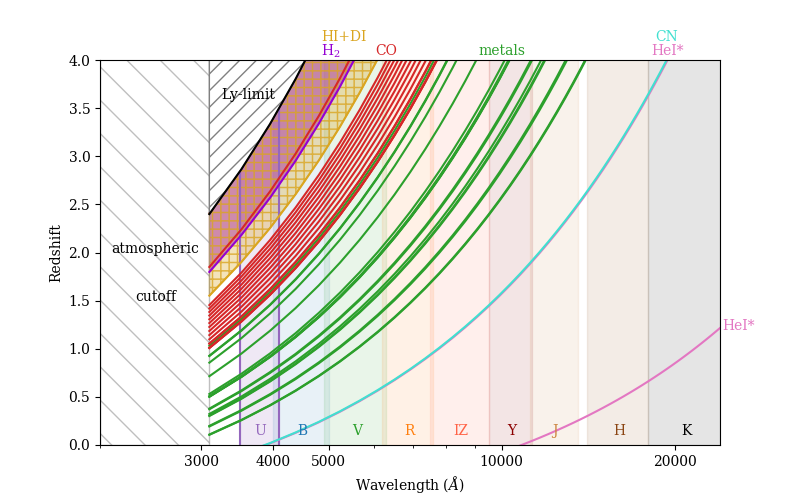}
\caption{Redshift-wavelength relation for the species of interest to the science cases discussed in the present work: metals for $\alpha$, H$_2$ for $\mu$, DI, HI and HeI* for BBN, CO, H$_2$ and CN for the CMB temperature, and HI for the redshift drift. The foreseen ANDES bands are also displayed. Bearing in mind that most (bright) quasars are at $z\leq3$, this shows that the U band is essential for ANDES to be a competitive fundamental physics probe in the 2030s.}\label{f:redshifts}
\end{figure*}

Last but not least, we must emphasize the the answer to the question of how competitive ANDES will be as a cosmology and fundamental physics probe, in the context of other ground and space facilities operating in the 2030s and beyond, depends almost entirely on two factors. The first is the availability of the U band, which is currently a goal; its need is scientifically motivated in several earlier sections of this work, and can also be visually  understood from Figure \ref{f:redshifts}. The second and equally important one is its ability to fully benefit from the ELT's larger collecting area (i.e., ensuring that we do not become systematics-limited). This implies further developments not only in terms of wavelength calibration but in the full data reduction and analysis pipeline. Admittedly some of the required developments are challenging (and possibly costly), but one must keep in mind that, in a rapidly developing and data-driven field like fundamental cosmology, something that is easy to do today is very unlikely to be scientifically competitive in ten years' time.

\backmatter

\bmhead{Acknowledgments}

This work was financed by Portuguese funds through FCT (Funda\c c\~ao para a Ci\^encia e a Tecnologia) in the framework of the project 2022.04048.PTDC (Phi in the Sky, DOI 10.54499/2022.04048.PTDC). CJM also acknowledges FCT and POCH/FSE (EC) support through Investigador FCT Contract 2021.01214.CEECIND/CP1658/CT0001.
RJC is funded by a Royal Society University Research Fellowship, and acknowledges support from STFC (ST/T000244/1).
MTM acknowledges the support of the Australian Research Council through Future Fellowship grant FT180100194.
JL acknowledges support by the Deutsche Forschungsgemeinschaft (DFG, German Research Foundation) under Germany’s Excellence Strategy – EXC 2121 ``Quantum Universe'' – 390833306.
TMS acknowledges the support from the SNF synergia grant CRSII5-193689 (BLUVES).
JIGH acknowledges financial support from the Spanish Ministry of Science and Innovation (MICINN) project PID2020-117493GB-I00.
CMJM is supported by an FCT fellowship, grant number 2023.03984.BD. 

\bmhead{Author contributions}
CJAPM: Coordination, writing of Sects. 1 and 7; RC: Writing of Sect. 2; JL: Writing of Sect. 5; MTM: Writing of Sect. 4; PN: Writing of Sect. 3; TMS: Writing of Sect. 6; CSA, SB, CMJM, MAFMS, SV: Help with simulations/forecasts; PDM, RM, AM, LO, AZ: ANDES Project Office support; JSA, SC, RGS, RSG, JIGH, NJN, CP: Comments/suggestions on the draft.

\bmhead{Data availability}
Data sharing is not applicable to this article as no datasets were generated or analysed during the current study.

\bmhead{Conflicts of interest}
The authors declare that they have no conflict of interest.


\bibliography{andeswg4}% common bib file

\end{document}